# Biquadratic exchange coupling effect on the magnetic properties of (Fe/Ti) multilayers

M. Yactayo*, H. S. Tarazona, O. Copie, J.-C. Rojas-Sánchez, J. Quispe-Marcatoma, C. V. Landauro

***Abstract*— This work explores the static and dynamic magnetic properties of weakly antiferromagnetically coupled Fe/Ti superlattices, emphasizing the link between magnetic behavior and structural characteristics. HRTEM and XRD analyses confirm alternating Fe and Ti layers with rough interfaces, especially in upper layers. Magnetic measurements reveal two-step hysteresis loops and a temperature- and thickness-dependent interlayer exchange coupling (IEC). A macrospin model incorporating bilinear and biquadratic coupling reproduces the experimental data and supports a phase diagram analysis identifying non-collinear configurations. The results underscore the impact of structural imperfections and highlight the crucial role of biquadratic exchange in Fe/Ti/Fe multilayers.**

## I. Introduction

Magnetic multilayer systems have attracted extensive interest due to their unique exchange interactions and potential applications in magnetic storage, spintronics, and sensing technologies [1,2]. Among these, Fe/Ti/Fe trilayers are especially compelling because of the weak antiferromagnetic coupling mediated by the Ti spacer—an element known for its low magnetic transparency [2,3]. Understanding the interplay between structural parameters and magnetic coupling is essential for optimizing the performance of such systems in device contexts [3-5]. In these heterostructures, bilinear exchange coupling typically governs the antiparallel alignment of adjacent ferromagnetic layers, giving rise to technological phenomena such as giant and tunneling magnetoresistance [2, 6-8]. Biquadratic exchange coupling, a higher-order interaction, is invoked to explain perpendicular (90°) orientations between magnetizations and plays a critical role in stabilizing complex spin textures in multilayers [9-11]. Studies have demonstrated that competition between bilinear and biquadratic coupling can dictate magnetic states and dynamic behavior, making these interactions key enablers for next-generation spintronic and multi-state memory devices [10,12,13]. The microscopic origin of biquadratic coupling is often attributed to interface imperfections such as roughness, spin-disordered regions in the spacer, or thickness fluctuations [9-11,14]. In Fe/Ti systems, interdiffusion at interfaces can form amorphous or compound layers that affect magnetic properties [5,14,15]. For instance, recent work on Nd-Fe-B/Ti/Fe heterostructures has revealed step-like hysteresis, indicative of antiferromagnetic alignment at ~2 nm Ti thickness, aligning with RKKY-type oscillatory coupling models [16]. Similarly, thermal treatments at Fe–Ti interfaces have shown structural evolution that significantly alters interlayer exchange [5]. These results reinforce the sensitivity of magnetic coupling to both interface chemistry and processing history, particularly in systems utilizing Ti spacers. Temperature-dependent investigations have further clarified how thermal energy modulates interlayer exchange strength. Experiments on $[Pt/Co]_3$/NiO/$[Co/Pt]_3$ multilayers [12] and other systems [13] have shown that temperature can drive reversible transitions between magnetic configurations, offering tunable control over bilinear and biquadratic terms. Structural characterization via X-ray reflectivity (XRR) and HRTEM has confirmed that substrate choice and deposition conditions influence layer roughness and morphology, thereby affecting the magnitude of magnetic coupling [5,17,18]. Furthermore, magnetically dead layers on the order of 1 nm have been observed in Fe-based stacks due to intermixing at interfaces, with thicknesses strongly tied to deposition and annealing parameters [15,19]. Against this backdrop, our study aims to correlate structural and magnetic properties in Fe/Ti/Fe trilayers to provide insight into their coupling behavior and temperature evolution—information crucial for tailoring multilayer performance in spintronic applications. Therefore, this work focuses on analyzing how the structure and morphology influence the magnetic behavior of Fe/Ti/Fe multilayers, providing relevant insights for future technological applications.

## II. Materials and Methods

Weakly exchange-coupled Fe/Ti superlattices were fabricated using the RF confocal magnetron sputtering technique. The multilayer structure is: Substrate/Ti(4.1 nm)/[Fe(2.3 nm)/Ti(1.8 nm)/Fe(2.3 nm)/Ti(4.1 nm)]$_n$, where N is the number of trilayer repeats and ranges from 1 to 6. Glass substrates were used for structural characterization, and Si(111) substrates (with native oxide) were used for magnetic measurements. The base pressure was $8\times10^{-7}$ Torr and the deposition occurred at room temperature and a working pressure of $4\times10^{-3}$ mTorr Ar gas with the substrate rotating 20 rpm to ensure uniform layer deposition. Structural and morphological characterization was performed using High-Resolution Transmission Electron Microscopy (HRTEM) and Scanning Transmission Electron Microscopy (STEM), including HAADF imaging and Electron Energy-Loss Spectroscopy (EELS) for elemental

* Research supported by VRIP-UNMSM, ERC CoG project MAGNETALLIEN grant ID 101086807 and ULTIMATE-I, Grant ID. 101007825.

M. Yactayo is with the Université de Lorraine, CNRS, Institut Jean Lamour – CNRS:UMR7198, France and Universidad Nacional Mayor de San Marcos – Peru (corresponding author: melissa-sonia.yactayo-yaranga@univ-lorraine.fr). H. S. Tarazona is with the Universidad Nacional Mayor de San Marcos – Peru.(e-mail: heisemberg.tarazona@unmsm.edu.pe) O. Copie is with the Université de Lorraine, CNRS, Institut Jean Lamour – CNRS:UMR7198, France (e-mail:olivier.copie@univ-lorraine.fr). J.-C. Rojas-Sánchez Université de Lorraine, CNRS, Institut Jean Lamour – CNRS:UMR7198, France (e-mail: juan-carlos.rojas-sanchez@univ-lorraine.fr). J. Quispe-Marcatoma Universidad Nacional Mayor de San Marcos – Peru.(e-mail:jquispem@unmsm.edu.pe) C. V. Landauro Universidad Nacional Mayor de San Marcos – Peru (e-mail: clandauros@unmsm.edu.pe).

analysis, Fig. 1. Crystallographic structure was examined using X-ray Diffraction (XRD). X-ray Reflectivity (XRR) was used to determine the total film thickness and multilayer periodicity, Fig 2. Static magnetic properties were measured using a SQUID magnetometer over a temperature range of 10 - 300 K, Fig. 3b-c. Both full major hysteresis loops and minor loops were measured to evaluate interlayer coupling. Dynamic properties were probed by Ferromagnetic Resonance (FMR) using a VNA-FMR setup, Fig. 3d. The samples (on Si substrates) were placed on a coplanar waveguide, and FMR absorption spectra were recorded at room temperature for frequencies 2-20 GHz with the static field applied in-plane (parallel to the sample plane. Field-swept FMR measurements at fixed frequencies allowed extraction of resonance field $\mu_0 H_r$ vs frequency dispersion relations.

## III. THEORETICAL MODELING

To interpret the magnetic measurements, we employed a macrospin model for Fe/Ti/Fe trilayers that includes both bilinear and biquadratic interlayer exchange couplings. Each Fe layer behaves as a single domain with magnetization vectors $M_1$ and $M_2$. A small effective in-plane uniaxial anisotropy was included in the model. The coordinate system is defined considering the multilayer film plane lying in the *xy*-plane and the normal (growth direction) along *z* (denoted by unit vector n̂), Fig. 3(a). The total free energy per unit area of one Fe/Ti/Fe trilayer (building block) can be written as:

$$E = t_1[-M_1 H \sin\theta_1 \cos(\alpha - \varphi_1) - K_{u_1} \sin^2\theta_1 \cos^2\varphi_1 - K_{eff_1} \cos^2\theta_1] + t_2[-M_2 H \sin\theta_2 \cos(\alpha - \varphi_2) - K_{u_2} \sin^2\theta_2 \cos^2\varphi_2 - K_{eff_2} \cos^2\theta_2] - J_{bl}[\sin\theta_1 \sin\theta_2 \cos(\varphi_1 - \varphi_2) + \cos\theta_1 \cos\theta_2] - J_{bq}[\sin\theta_1 \sin\theta_2 \cos(\varphi_1 - \varphi_2) + \cos\theta_1 \cos\theta_2]^2 \quad (1)$$

where $t_i$ is the thickness of the i-th Fe layer, $M_i$ its saturation magnetization, H the applied magnetic field magnitude, and α the in-plane angle of H (with respect to x axis). $Ku_i$ represents an effective in-plane uniaxial anisotropy (with easy axis along $\varphi = 0$ and $H_{u_i} = 2K_{u_i}/M_i$ as the in-plane anisotropy field of layer i), and $K_{eff_i}$ represents an effective perpendicular (out-of-plane) anisotropy (which includes demagnetizing effects for thin films). The indices i=1,2 correspond to the bottom and top Fe layers of a single trilayer. The last two terms of Eq. 1 describe the interlayer exchange coupling energy, with $J_{bl}$ and $J_{bq}$ the bilinear and biquadratic coupling constants, respectively. For the $J_{bq}$ term, a positive $J_{bq}$ favors collinear alignments (either parallel or antiparallel, since the term involves $\cos^2(\varphi_1 - \varphi_2)$, whereas a negative $J_{bq}$ favors a perpendicular (90°) orientation between $M_1$ and $M_2$. Prior theoretical work has shown how competing $J_{bl}$ and $J_{bq}$ can lead to non-collinear ground states [10,20]. The inclusion of a biquadratic term in our model is further supported by the presence of rough interfaces and possible magnetically dead (uncompensated) interfacial regions, as evidenced by our structural analysis, since such imperfections can give rise to a biquadratic coupling component. We implemented a numerical routine to find the equilibrium $\varphi_1(H)$, $\varphi_2(H)$ by gradually stepping the field and using a relaxation algorithm to minimize *E* at each step. This method allows simulation of static and dynamic magnetic properties such as hysteresis loops and ferromagnetic resonance for comparison with experiment. Further details of the loop-fitting procedure and stability criteria can be found in Ref. [21].

Thus, for the modeling of hysteresis loops, the total multilayer magnetization (projected along the field direction) is computed as:

$$M_{tot}(H) = [t_1 M_1 \cos(\alpha - \varphi_1) + t_2 M_2 \cos(\alpha - \varphi_2)]/(t_1 + t_2).$$

For the modeling of the FMR behavior, we use the Landau-Lifshitz equation (without damping) for small oscillations about equilibrium; i.e.:

$$\partial M_i(t)/\partial t = -\gamma[M_i(t) \times H_{eff,i}]$$

where γ is the gyromagnetic ratio and $H_{eff,i} = -(1/(t_i M_i))(\partial E/\partial M_i)$ is the effective field on layer i. For small precessional motions, one can linearize about the equilibrium orientation (given by φ) using the Smit-Beljers approach [22]. This leads to a characteristic equation for the resonance modes. In our case (two coupled layers), there are two modes (acoustic and optical). The resonance condition can be written as a quartic in the angular frequency ω, which, when expanded, yields:

$$\omega^4 - b\omega^2 + c = 0$$

with coefficients b and c expressed in terms of the field H, the anisotropy fields $Hu_i$, the effective perpendicular fields $H_{effi}$, and the coupling constants $J_{bl}$, $J_{bq}$

$$b = -\frac{\gamma_2^2}{t_2^2 M_2^2}\left(E_{\theta_2\theta_2} E_{\phi_2\phi_2}\right) - \frac{\gamma_1^2}{t_1^2 M_1^2}\left(E_{\phi_1\phi_1} E_{\theta_1\theta_1}\right) - \frac{2\gamma_1\gamma_2}{t_1 M_1 t_2 M_2}\left(E_{\phi_1\phi_2} E_{\theta_1\theta_2}\right)$$

$$c = \frac{\gamma_2^2}{t_2^2 M_2^2}\frac{\gamma_1^2}{t_1^2 M_1^2}\left({E^2}_{\phi_1\phi_2} - E_{\phi_2\phi_2} E_{\phi_1\phi_1}\right)\left({E^2}_{\theta_1\theta_2} - E_{\theta_1\theta_1} E_{\theta_2\theta_2}\right)$$

More details of the method to obtain these expressions can be found in Ref. [22]. The model parameters ($t_1$, $t_2$, $M_1$, $M_2$, $H_{u_1}$, $H_{u_2}$, $H_{eff_1}$, $H_{eff_2}$, $J_{bl}$, $J_{bq}$) were adjusted to achieve the best agreement with both the static M-H loops and the FMR spectra for each sample.

## IV. RESULTS AND DISCUSSION

### A. *Structural analysis*

The structural analysis using HRTEM revealed rough interfaces, with this roughness being more noticeable in the top layers (Fig. 1). Experimental layer thicknesses were determined for instance 2.7 nm, 3.7 nm and 5.8 nm for iron and the thinner and thicker titanium layers, respectively. From electron diffraction patterns (Inset Fig. 1), BCC α-Fe was identified, with preference orientation in [110]. It was corroborated by XRD. No other phases were recognized through these techniques, suggesting an amorphous Ti phase [4]. Fe lattice parameter showed a slight distortion (a 0.73% deviation from the bulk value), indicating the presence of strain in the crystalline structure, which is a result of the growth technique. Elemental analysis using HAADF and EELS identified iron (Fe-$L_{2,3}$), (Ti-$L_{2,3}$), and oxygen (O-K). EELS maps showing oxygen predominantly overlaps with the titanium layers, due to Ti has strong affinity with oxygen (likely contaminated after Focused ion beam etching lamella preparation). Fig. 1, revealed diffuse interfaces due to slight intermixing between Fe, Ti, and O, or structural imperfection defects [3]. This interdiffusion of elements at the interfaces, result in the formation of amorphous alloys or an intermetallic Fe/Ti phase [2-5]. Atomic resolution images confirmed that the interfaces are clearly rough and that this roughness, resulting from atomic mixing, increases with the number of layers. This structural analysis is relevant because the magnetic and structural properties of the material are strongly linked. The X-ray reflectivity pattern for the N=6 sample revealed that the glass substrate yielded a lower oscillation amplitude and a faster decrease in reflected X-rays compared to the silicon substrate, indicating higher roughness in the films grown on glass [17]. By performing a linear fit of the (peak order)$^2$ versus $\sin^2\theta$, where θ is the angle at which the interference occurs [18,25], the multilayer period thickness ($t_{stack}$) was estimated to be 15.1 nm (Fig. 2(b)), and the total superlattice thickness $\Lambda_{N=6}$ was determined to be 92.2 nm (Fig. 2(c)). The values obtained by XRR agree with the experimental thickness results obtained by HRTEM.

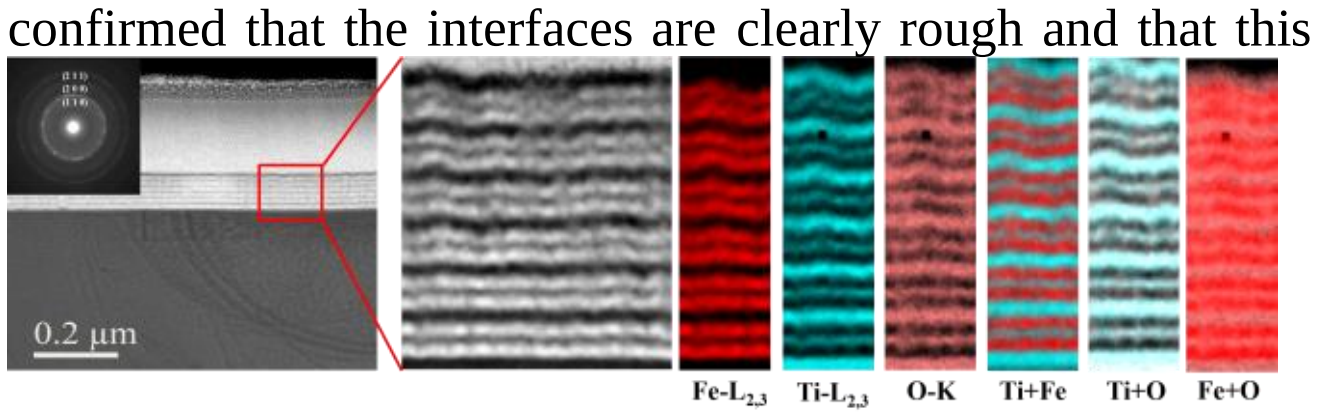

Figure 1. Cross-sectional HAADF-STEM image of the multilayer (N = 6). The inset shows the electron diffraction pattern, with reflections indexed to bcc Fe and a [110] texture. The red rectangle marks the area analyzed by EELS. EELS maps of the elements identified are shown on the right, such as Fe-$L_{2,3}$ (red), Ti-$L_{2,3}$ (cyan), O-K (pink), and their overlap.

## B. *Magnetic properties*

All samples exhibit in-plane magnetic hysteresis loops characteristic of two weakly coupled ferromagnetic layers. For the simplest trilayer (N=1) measured at 300 K, the major hysteresis loop exhibits a two-step switching behavior: one Fe layer reverses its magnetization at a lower coercive field ($H_{c_1}$), while the other switches at a higher coercive field ($H_{c_2}$), as it is shown in Fig. 3(b). This indicates antiferromagnetic interlayer coupling whereby the layers energetically favor an antiparallel alignment over a range of applied fields.

The presence of a small but finite remanent magnetization suggests a canted or uncompensated antiparallel state, likely arising from slight differences in layer magnetizations due to thickness variations or magnetically dead interfacial layers, resulting effectively in a ferrimagnetic-like configuration. To quantify the saturation magnetization and assess the presence of magnetically dead layers, we plotted the total magnetic moment versus total Fe thickness, obtaining a linear fit whose extrapolation to zero moment yields an estimated dead layer thickness of approximately 1.1 nm per Fe/Ti interface.

The effective saturation magnetization of the active Fe layers was found to be slightly reduced (~1.3–1.4 × $10^6$ A/m) compared to bulk Fe (~1.7 × $10^6$ A/m), consistent with interfacial interdiffusion and oxidation.

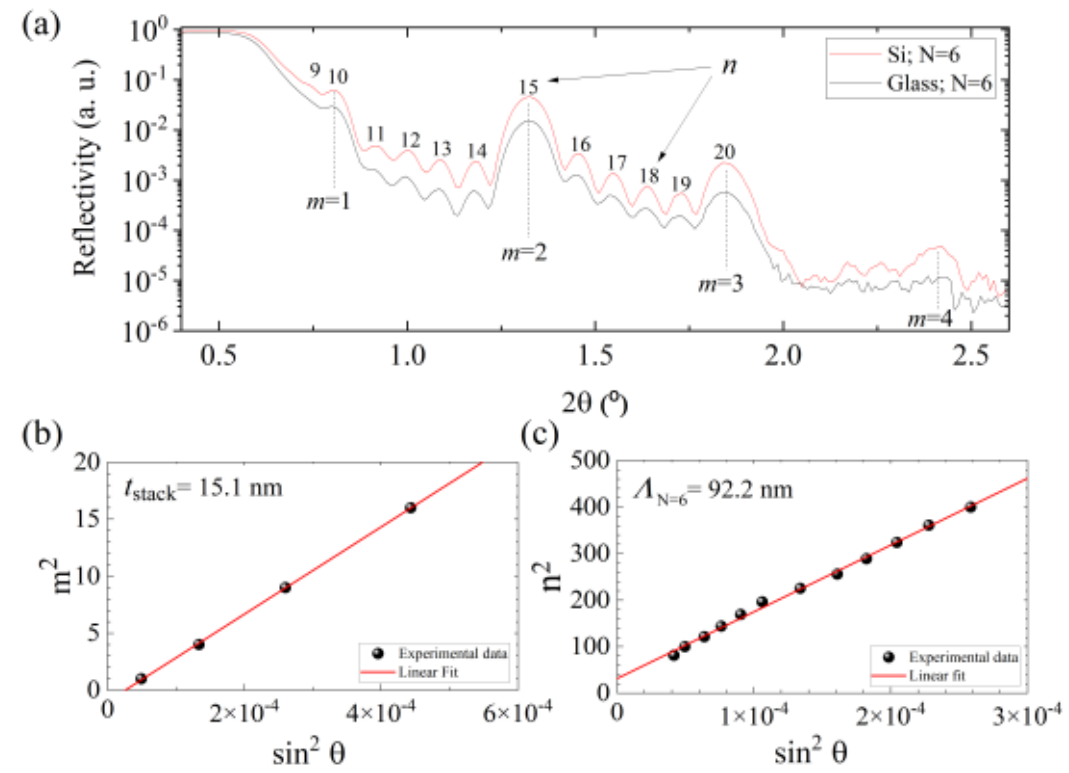

Figure 2. (a) Normalized reflectivity profile of sample N=6 grown on glass and silicon substrates, notice that the first present higher roughness than the second substrate. Letter *m* represents the peaks order of the superlattice peaks and letter *n* the Kiessig fringes order. Linear fit of (peak order)$^2$ vs $\sin^2\theta$ to obtain the b) multilayer period thickness ($t_{stack}$=15.1 nm), and the total superlattice thickness for samples with higher numbers of repetition, $\Lambda_{N=6}$=92.2 nm.

Minor hysteresis loop (Fig. 3(c)) measurements revealed a positive minor loop shift field ($H_{mls}$) corresponding to antiferromagnetic coupling, with the effective interlayer exchange coupling strength ($J_{IEC}$) on the order of a few tenths of mJ/m², increasing with the number of repetitions N up to ~0.5 mJ/m² at N=6. Despite increased interface roughness at higher N, coupling strength was enhanced, possibly due to improved quality at internal interfaces or cumulative dipolar interactions among multilayers. Overall, the static magnetometry confirms a baseline weak antiferromagnetic coupling robust enough to manifest distinct two-step switching in these sputtered Fe/Ti multilayers.

We additionally carried out ferromagnetic resonance (FMR) measurements to investigate the magnetization dynamics in the Fe/Ti/Fe trilayers and to independently validate the exchange coupling parameters extracted from static measurements. As shown in Fig. 3(d), the theoretical model exhibits excellent agreement with the experimental FMR data for the N = 1 sample, underscoring the success of the model in capturing both the static and dynamic magnetic behavior of the system.

### *B.1. Analytical modeling of hysteresis*

Figure 4 presents experimental and simulated hysteresis loops for the N = 1 sample at various temperatures, demonstrating excellent agreement between model and measurement. At 300 K, a distinct two-step loop is observed, indicative of antiparallel coupling at intermediate fields. As temperature is reduced to 50 K, this feature collapses into a nearly single-step reversal, suggesting weakened coupling or a shift toward collinear rotation. Remarkably, at 10 K the two-step behavior reappears prominently, consistent with enhanced antiferromagnetic coupling at low temperature.

These trends are accurately captured by the model through temperature-dependent adjustments of the bilinear ($J_{bl}$) and biquadratic ($J_{bq}$) exchange constants (see Table 1).

The inclusion of $J_{bq}$ is essential to reproduce the temperature-driven evolution in loop shape—particularly the suppression and re-emergence of the two-step signature.

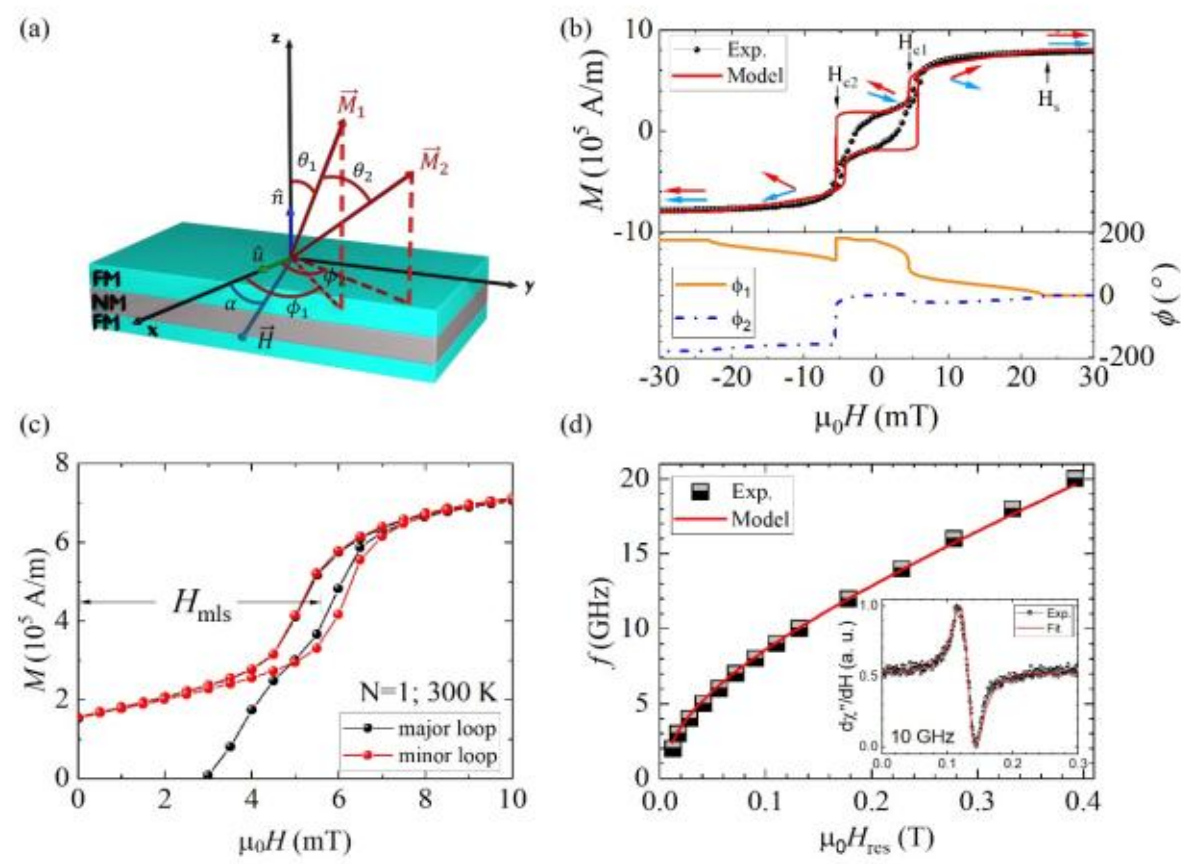

Figure 3. (a) Schematic diagram of the coordinate system and magnetization vectors used in the theoretical model. (b) Experimental and simulated magnetization hysteresis loops (top) and angular evolution of magnetizations (bottom) at 300 K. (c) Major and minor hysteresis loops showing the determination of the minor loop shift field $H_{mls}$. (d) Ferromagnetic resonance (FMR) frequency-field dependence for the sample with N=1; experimental data and model fit are shown. Inset: corresponding raw FMR signal at 10 GHz.

In the simulations, the Fe layer thicknesses were fixed at $t_1$ = 1.9 nm and $t_2$ = 2.1 nm, slightly below the nominal values to account for interfacial dead layers (~0.4 nm per interface). Saturation magnetizations $M_1$ and $M_2$ were treated independently, with $M_2$ consistently larger, likely due to the free surface of the top Fe layer experiencing less structural disorder than the Ti-seeded bottom layer. Both $M_1$ and $M_2$ decrease with increasing temperature, as expected, although a subtle anomaly was observed around 100–150 K, where both showed a temporary increase—possibly hinting at a compensation or transitional effect between layers. At 10 K, the extracted $J_{bl}$ is strongly negative (antiferromagnetic), while the positive $J_{bq}$ stabilizes collinear states, reinforcing the antiparallel configuration. These findings underscore the role of competing exchange interactions and asymmetry between layers in shaping the temperature-dependent reversal mechanisms.

TABLE 1. Parameters used to fit experimental hysteresis loops of the N=1 at various temperatures T.

| T (K) | $H_{u1}$ (mT) | $H_{u2}$ (mT) | $M_1$ ($10^5$ A/m) | $M_2$ ($10^5$ A/m) | $J_{bl}$ (mJ/m$^2$) | $J_{bq}$ (mJ/m$^2$) |
|---|---|---|---|---|---|---|
| 10 | 38 | 42 | 5.5 | 11.7 | -0.980 | 0.560 |
| 50 | 22 | 22 | 6.5 | 11.3 | -0.120 | 0.050 |
| 100 | 12 | 12.5 | 7.8 | 10.0 | -0.015 | -0.025 |
| 150 | 19 | 20 | 7.8 | 10.2 | -0.092 | -0.100 |
| 200 | 13.5 | 15 | 7.2 | 10.3 | -0.094 | -0.140 |
| 250 | 10 | 11 | 7.2 | 9.7 | -0.094 | -0.120 |
| 300 | 7 | 8.5 | 6.7 | 9.4 | -0.087 | -0.072 |

Thus at 10 K the ground state is antiparallel (as evidenced by the two-step loop with nearly zero net moment between steps). As temperature increases to 50 K, both |$J_{bl}$| and $J_{bq}$ drop dramatically in magnitude (Table 1 shows $J_{bl}$ = -0.12 mJ/m$^2$ and $J_{bq}$ = +0.05 mJ/m$^2$ at 50 K, an order of magnitude reduction). A negative biquadratic term implies that a 90° alignment of magnetizations can be preferred in zero field (non-collinear state). In our model, a negative $J_{bq}$ at high T helps to reproduce the fact that the mid-section of the hysteresis loop had a slanted shape approaching zero magnetization more gradually, consistent with a canted configuration rather than a fully antiparallel one.

The temperature dependence of both coupling constants is plotted in Fig. 5. It is confirmed that both |$J_{bl}$| and |$J_{bq}$| decrease upon increasing T due to thermal agitation which weaken the effective coupling (through magnon excitation and averaging out of spin correlations). Notably, $J_{bq}$ crosses from positive to negative around 75–100 K. The physical interpretation could be that at low T, interface imperfections like roughness or "loose spins" contribute to a $J_{bq}$ that is positive (e.g., the biquadratic coupling from fluctuations in spacer thickness tends to be positive according to Slonczewski's model [23]). Hence, the model provides an acceptable phenomenological simulation of our data, effectively describing the observations. Nevertheless, the physical mechanism behind the sign change remains unclear, demanding future study which is outside the scope of this investigation.

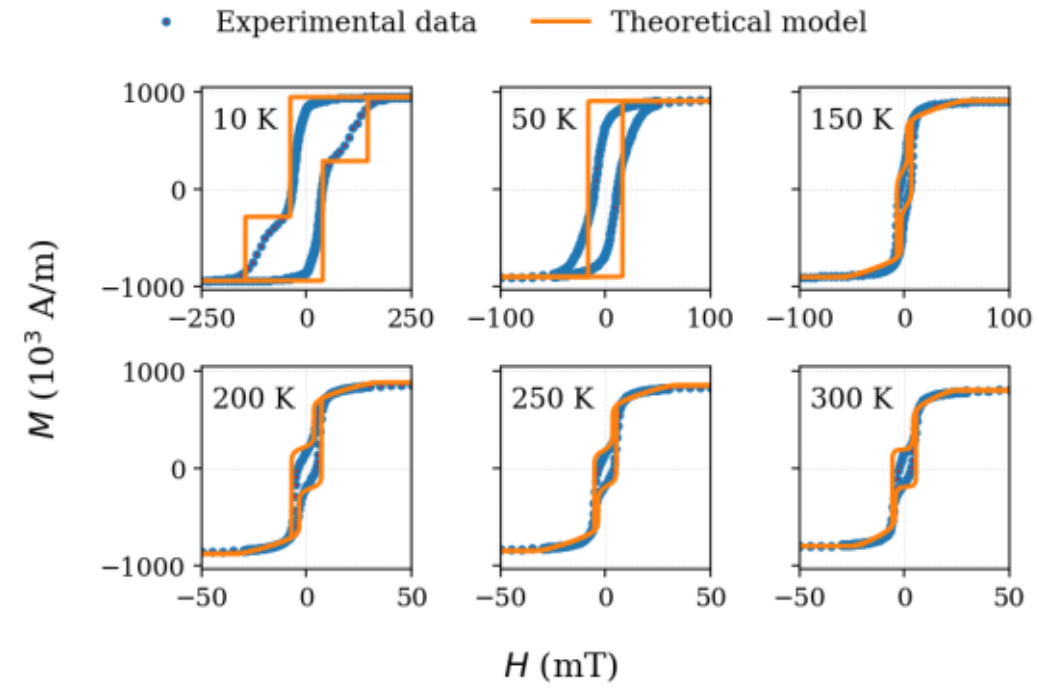

Figure 4. Hysteresis loops for N=1 (Fe/Ti/Fe trilayer on Si) at different temperatures (10 K, 50 K, 150 K, and 300 K). Open symbols are the experimental data; blue solid lines are simulated loops.

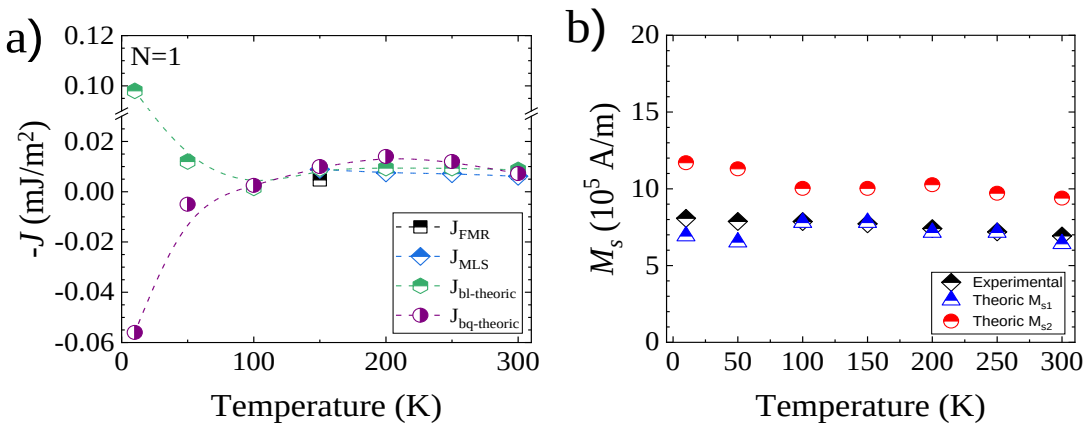

Figure 5. (a) Temperature dependence of the interlayer exchange coupling constants for the Fe/Ti/Fe trilayer with N=1. The plot shows the experimental bilinear coupling extracted from minor hysteresis loops ($J_{MLS}$) and FRM data ($J_{FMR}$), as well as the fitted theoretical values of bilinear ($J_{bl}$) and biquadratic ($J_{bq}$) coupling constants. (b) Temperature dependence of the net magnetization. Red open symbols correspond to experimental data, while blue and violet symbols represent theoretical saturation.

### *C. Phase diagram of $J_{bl}$ and $J_{bq}$*

To further elucidate the magnetic configuration tendencies of our coupled Fe/Ti/Fe system, we constructed an analytical phase diagram based on the zero-field energy minimization of the coupled layers. Following the formalism of A. Layadi [24], we derived the equilibrium conditions ($\partial E/\partial \varphi_i = 0$) and applied the stability criterion, wherein the determinant of the Hessian matrix vanishes, to obtain an analytical expression for the phase boundary separating distinct magnetization configurations:

$$J_{bl} + 2J_{bq} = -\frac{t_1 M_1 t_2 M_2 (H_s + H_{u1})(H_s + H_{u2})}{t_1 M_1 (H_s + H_{u1}) + t_2 M_2 (H_s + H_{u2})} \quad (2)$$

This analytical result matches well with the numerical phase diagram shown in Fig. 6, obtained by globally minimizing the total energy using a differential evolution algorithm for a range of bilinear ($J_{bl}$) and biquadratic ($J_{bq}$) coupling values. The diagram maps the angular difference $\Delta\varphi$ between the magnetizations of the Fe layers and reveals three distinct regions: an antiferromagnetic (AF) state with antiparallel alignment (↑↓), a non-collinear (NC) state with perpendicular orientation, and, beyond our experimental range, a parallel (↑↑) state for large positive $J_{bl}$ and $J_{bq}$. The overlaid data points in Fig. 6 illustrate how $J_{bl}$ and $J_{bq}$ evolve with temperature for N = 1, transitioning from an AF region at low T to a NC region at high T—consistent with the experimentally observed changes in hysteresis loop shapes.

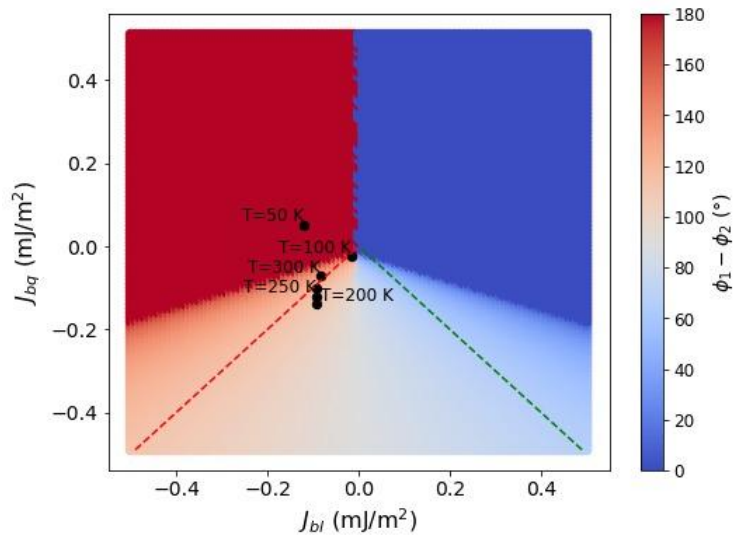

Figure 6. Theoretical phase diagram showing the preferred relative orientation of the two Fe layer magnetizations as a function of the bilinear coupling ($J_{bl}$) and biquadratic coupling ($J_{bq}$) (at zero applied field).

The phase boundaries indicate where two configurations become energetically degenerate, and their precise positions depend on the in-plane anisotropy fields $H_{u1}$ and $H_{u2}$, as described in Eq. (2). Specifically, the AF region is stabilized when $|J_{bl}|$ dominates over $J_{bq}$, while a sufficiently negative $J_{bq}$ can promote a perpendicular alignment even when $J_{bl}$ remains weakly negative. Dashed lines in Fig. 6 mark the conditions $J_{bl} = J_{bq}$ and $J_{bl} = -J_{bq}$, highlighting the symmetry in phase space. Although an external magnetic field would shift these boundaries and potentially stabilize a canted state, the zero-field phase diagram serves as a powerful conceptual framework. These findings reinforce the view that structural imperfections and thermal fluctuations influence the effective coupling parameters, driving the system through collinear to non-collinear magnetic states. Such transitions, seen in other multilayer systems and relevant for spintronic applications like multi-state MRAM, demonstrate that Fe/Ti trilayers reside near the boundary between these magnetic regimes.

*D. Dynamic magnetic properties (FMR)*

Ferromagnetic resonance spectra recorded from 2 to 20 GHz for samples with N = 1 to 6 reveal a clear frequency-dependent resonance shift, consistent with Kittel behavior, as it is shown in Fig. 7. In most samples, two coupled resonance modes were observed—acoustic and optic—although they often overlapped due to the small difference in magnetic properties between the two Fe layers.

For N = 6, weak mode splitting was evident at low frequencies, indicating the presence of interlayer coupling. These features were captured by fitting the FMR dispersion curves (resonance field vs. frequency) using a coupled macrospin model, treating each trilayer repeat as an isolated entity owing to the thick Ti spacer (4.1 nm) between repeats, which effectively suppresses inter-block magnetic interactions. Fig. 7a shows the results at 300 K confirming the presence of strong in-plane anisotropy (negative $H_{eff}$ values), in the top Fe layers ($H_{eff2} \approx -0.8$ to $-1$ T), while bottom layers exhibit weaker anisotropy, potentially due to Ti-seeded interface effects.

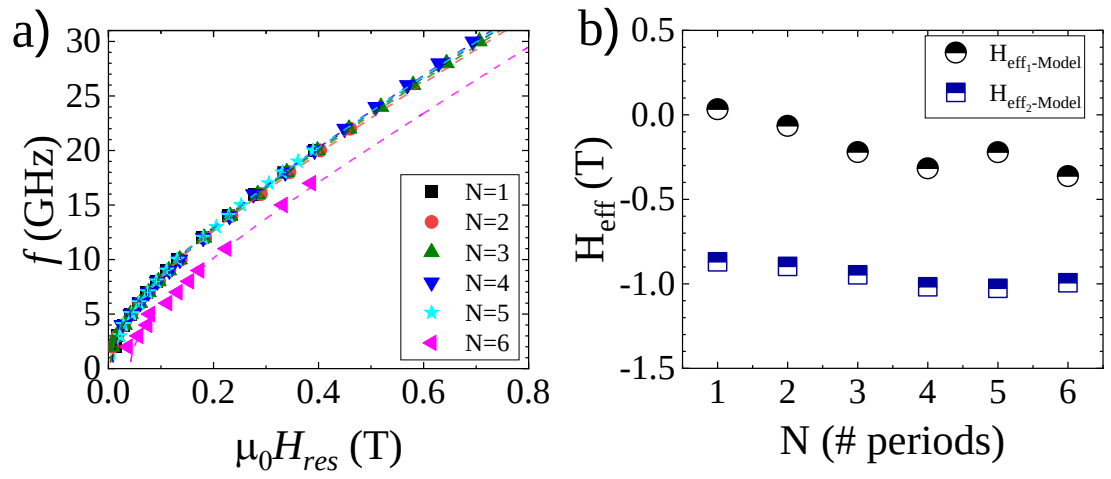

Figure 7. (a) Frequency dependence of the ferromagnetic resonance field ($\mu_0 H_{res}$) for Fe/Ti/Fe multilayers with N = 1 to 6 repeats. Experimental data (symbols) and model fits (dashed lines) show good agreement across all samples, with slight variation in curvature indicating differences in effective anisotropy. (b) Extracted effective perpendicular anisotropy fields ($H_{eff}$) for the bottom (circles) and top (squares) Fe layers as a function of the number of trilayer periods. Both layers exhibit in-plane easy-axis character (negative $H_{eff}$), with the top layer showing a consistently stronger easy-plane tendency.

TABLE 2. Parameters used to fit experimental FMR data for N=1,2,3,4,5 and 6.

| **N** | **$H_{eff1}$ (mT)** | **$H_{eff2}$ (mT)** | **$H_{u1}$ (mT)** | **$H_{u2}$ (mT)** | **$M_1$ ($10^5$ A/m)** | **$M_2$ ($10^5$ A/m)** | **$J_{bl}$ (mJ/m²)** | **$J_{bq}$ (mJ/m²)** |
|---|---|---|---|---|---|---|---|---|
| 1 | 0.03 | -0.87 | 7 | 8.5 | 6.4 | 9.4 | -0.087 | -0.072 |
| 2 | -0.07 | -0.90 | 7 | 8.5 | 10.4 | 12.8 | -0.22 | -0.16 |
| 3 | -0.22 | -0.95 | 7 | 8.0 | 12.1 | 14.3 | -0.30 | -0.25 |
| 4 | -0.32 | -1.02 | 7 | 8.0 | 10.0 | 11.6 | -0.40 | -0.30 |
| 5 | -0.22 | -1.03 | 7 | 8.0 | 9.4 | 10.4 | -0.35 | -0.27 |
| 6 | -0.36 | -1.00 | 7 | 8.0 | 11.8 | 12.8 | -0.32 | -0.27 |

The best-fit parameters from the frequency – resonance field are listed in Table 2. The in-plane anisotropy fields $H_u$, obtained from SQUID measurements, were weak and held constant. $H_u$ likely arises from both residual stress and a slight oblique incidence component inherent to the confocal sputtering geometry. The saturation magnetizations $M_1$ and $M_2$ shown in Table 2 were in good agreement with static magnetometry. Notably, the coupling constants $J_{bl}$ and $J_{bq}$ extracted from FMR also followed the same trend observed in minor-loop SQUID analysis: they increased in magnitude with N up to N = 4 ($J_{bl} \approx -0.40$, $J_{bq} \approx -0.30$, in mJ/m²) and slightly decreased beyond that. This suggests optimal coupling occurs in mid-range repeat numbers, likely due to improved interface quality or cumulative dipolar effects from adjacent trilayers. Importantly, including biquadratic coupling ($J_{bq}$) in the FMR model was essential to accurately capture the mode behavior; purely bilinear models failed to reproduce the observed resonance splitting. The consistency between dynamic and static fits highlights the robustness of the extracted parameters and supports the physical significance of $J_{bq}$ in determining not only equilibrium states but also spin-wave dynamics. These findings underscore the necessity of accounting for both bilinear and biquadratic exchange in multilayer systems, especially when analyzing frequency-dependent behavior in weakly coupled ferromagnetic heterostructures such as Fe/Ti.

## V. CONCLUSION

In conclusion, we demonstrate that even in a system where the bilinear IEC is weak, the presence of structural imperfections gives rise to a biquadratic exchange component that significantly alters the magnetic response of the Fe/Ti superlattice. To achieve this, we use repetitions of three layers $(Fe/Ti/Fe)_N$ with structural, magnetic (static and dynamic) characterizations, together with ad hoc macroscopic modeling of our experimental results, along with temperature dependance. For device applications, understanding and controlling this biquadratic coupling could be crucial, as it may affect the magnetization reversal process and the stability of antiparallel configurations. Our findings are relevant not only to Fe/Ti, but to other magnetic multilayers where interface quality is limited; in such cases, accounting for biquadratic coupling is necessary for accurate modeling. Future work could explore how to engineer $J_{bq}$ (for instance, by inserting ultrathin layers of impurities or adjusting roughness) to achieve desired non-collinear magnetic configurations, which are of interest for advanced spintronic devices. Additionally, temperature-dependent studies of coupling, as presented here, offer insight into the thermodynamics of exchange interactions and could guide the design of materials that exploit thermal switching of coupling.

## ACKNOWLEDGMENT

H.S. T. thanks the Peruvian Doctoral Scholarship Program of CIENCIACTIVA (CONCYTEC) for financial support under Grand No. 218-2014-FONDECYT. An Research supported by VRIP-UNMSM through the project ID B1813002. This work was partially funded by the ERC CoG project MAGNETALLIEN grant ID 101086807, the EU-H2020-RISE project Ultra Thin Magneto Thermal Sensoring ULTIMATE-I (Grant ID. 101007825), and the French National Research Agency (ANR) through the project “Lorraine Université d’Excellence” reference ANR-15-IDEX-04-LUE.